\documentclass[fleqn,12pt]{wlscirep_}
\title{Tensile strain-induced softening of iron at high temperature}

\usepackage{float}
\usepackage{gensymb}
\usepackage{setspace}
\author[1,*]{Xiaoqing Li}
\author[1,*]{Stephan Sch\"onecker}
\author[2]{Eszter Simon}
\author[3]{Lars Bergqvist}
\author[1,4]{Hualei Zhang}
\author[2,5]{L\'{a}szl\'{o} Szunyogh}
\author[6,*]{Jijun Zhao}
\author[1,7]{B\"{o}rje Johansson}
\author[1,7,8]{Levente Vitos}
\affil[1]{Department of Materials Science and Engineering, KTH - Royal Institute of Technology, 10044, Stockholm, Sweden}
\affil[2]{Department of Theoretical Physics, Budapest University of Technology and Economics, Budafoki \'ut 8., HU-1111, Budapest, Hungary}
\affil[3]{Department of Materials and Nano Physics, KTH Royal Institute of Technology, Electrum 229, SE-16440, Kista, Sweden}
\affil[4]{Center of Microstructure Science, Frontier Institute of Science and Technology, Xi'an Jiaotong University, 710054, Xi'an, China}
\affil[5]{MTA-BME Condensed Matter Research Group, Budafoki \'ut 8., HU-1111, Budapest, Hungary}
\affil[6]{Key Laboratory of Materials Modification by Laser, Ion and Electron Beams (Dalian University of Technology), Ministry of Education, 116024, Dalian, China}
\affil[7]{Department of Physics and Astronomy, Division of Materials Theory, Uppsala University, Box 516, SE-75120, Uppsala, Sweden}
\affil[8]{Research Institute for Solid State Physics and Optics, Wigner Research Center for Physics, P.O. Box 49, HU-1525, Budapest, Hungary}
\affil[*]{xiaoqli@kth.se}
\affil[*]{stesch@kth.se}
\affil[*]{zhaojj@dlut.edu.cn}

\begin{abstract}

In weakly ferromagnetic materials, already small changes in the atomic configuration triggered by temperature or chemistry can alter the magnetic interactions responsible for the non-random atomic-spin orientation. Different magnetic states, in turn, can give rise to substantially different macroscopic properties. A classical example is iron, which exhibits a great variety of properties as one gradually removes the magnetic long-range order by raising the temperature towards and beyond its Curie point of $T_{\text{C}}^{0}=1043$\,K. Using first-principles theory, here we demonstrate that uniaxial tensile strain can also destabilize the magnetic order in iron and eventually lead to a ferromagnetic to paramagnetic transition at temperatures far below $T_{\text{C}}^{0}$. In consequence, the intrinsic strength of the ideal single-crystal body-centered cubic iron dramatically weakens above a critical temperature of $\sim 500$\,K. The discovered strain-induced magneto-mechanical softening provides a plausible atomic-level mechanism behind the observed drop of the measured strength of Fe whiskers around $300-500$\,K. Alloying additions which have the capability to partially restore the magnetic order in the strained Fe lattice, push the critical temperature for the strength-softening scenario towards the magnetic transition temperature of the undeformed lattice. This can result in a surprisingly large alloying-driven strengthening effect at high temperature as illustrated here in the case of Fe-Co alloy.
\end{abstract}

\begin{document}
\maketitle
\newpage
Iron is one of the most abundant elements in the Milky Way and constitutes the major component of earth's core~\cite{Hemley:2001}. Processed iron is the main ingredient in steels and other ferrous materials, whose technology is a central pillar in today's industrial world. Beyond the obvious practical interests, the physical properties of Fe have attracted great attention in many fields of sciences. Probably the most conspicuous phenomenon is the marvellous coupling between magnetism and mechanical properties. Although recognized long time ago~\cite{Zener:1955}, it is only recently that atomic-level tools can give insight into the mechanisms behind the magneto-mechanical effects in Fe and its alloys.
Magnetic order in solid iron emerges with the formation of a crystalline lattice. The stable body-centered cubic (bcc) phase ($\alpha$-Fe) is characterized by long-range ferromagnetic (FM) order below $T_{\text{C}}^{0}$. The face-centered cubic (fcc) structure appears in the phase diagram at temperatures above 1189\,K in the paramagnetic (PM) state ($\gamma$-Fe). At low temperature, metastable fcc Fe has a complex antiferromagnetic state associated with a small tetragonal deformation~\cite{Tsunoda:2007,Meyerheim:2009}. This diversity corroborates the strong interplay between structural and magnetic degrees of freedom in Fe.

The weak FM nature of $\alpha$-Fe makes its magnetic interactions, and hence its characteristic properties, susceptible to chemical perturbations by, e.g., transition metal impurities. The effect is reflected in the well-known Slater-Pauling curves for the magnetic moment and Curie temperature of dilute Fe-based binaries~\cite{Takahasi:2007,Acet:2010}. The magneto-elastic coupling is manifested in the observed softening of the elastic parameters near the magnetic phase transition in bcc Fe~\cite{Dever:1972}.
The role of magnetism in the phase stability of hydrostatically pressurized bcc Fe has also been underlined~\cite{Soderlind:1996,Neumann:2004}. However, explorations of the interplay among magnetism, lattice strain, and micro-mechanical properties are missing hitherto.

Using computer simulations, we reveal the impact of the magneto-structural coupling on the temperature-dependent mechanical strength of bcc Fe under large anisotropic strain.
We apply uniaxial tensile load to a perfect bcc Fe single-crystal and investigate the non-linear stress-strain response up to the point of mechanical failure of the lattice (the maximum stress that a perfect lattice can sustain under homogeneous strain).
The stress at the failure point under tension defines the ideal tensile strength (ITS, denoted by $\sigma_\text{m}$), which is a fundamental intrinsic mechanical parameter of theoretical and practical importance~\cite{jiang:2013,Cerny:2013,Seung:2001,Li:2007,Thomson:1986,Jokl:1980}. We model the temperature effects in terms of a first-principles theoretical approach by taking into account contributions arising from phononic, electronic and magnetic degrees of freedom. The simulation details can be found in the method section.

\section*{Results and Discussions}
\begin{figure}[thb]
\begin{center}
\resizebox{0.7\columnwidth}{!}{\includegraphics[clip]{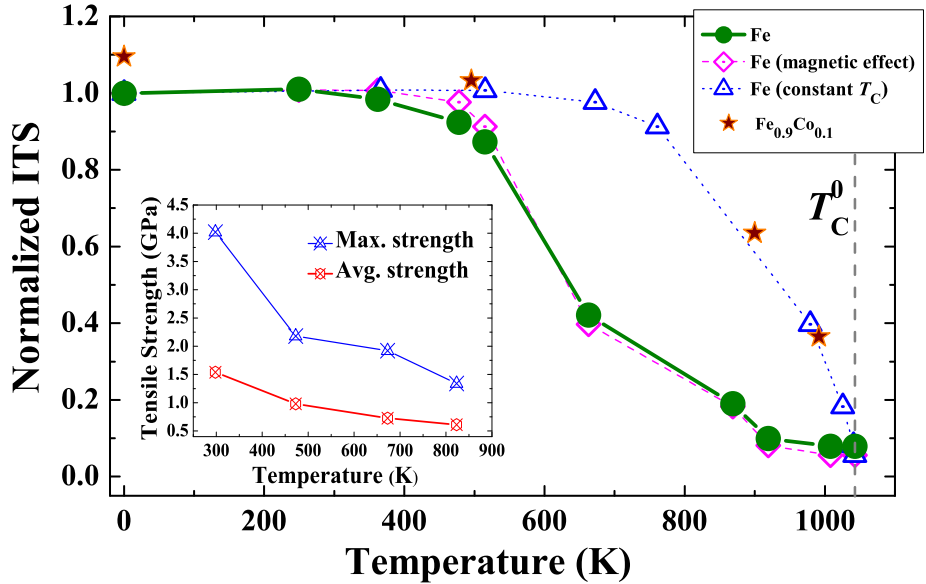}}
\caption{\label{fig:ITS}The ITS of bcc Fe in tension along the $[001]$ direction as a function of temperature ($\sigma_\text{m}(T)$, solid line and circles); the ITS taking into account only magnetic disorder (dashed line, open diamonds); the ITS taking into account magnetic disorder and neglecting the change of $T_\text{C}$ with structural deformation ($\sigma^c_\text{m}$, dotted line, open triangles). The Fe$_{0.9}$Co$_{0.1}$ alloy (stars) possesses a slightly larger ITS at 0\,K, but compared to pure Fe the ITS drastically increases at high temperatures. All data are normalized to the ITS of Fe at 0\,K (12.6\,GPa). (Inset) Experimentally determined tensile strength of $[001]$ oriented Fe whiskers as a function of temperature from Ref.~\cite{Brenner:1965}. The two data sets give the average tensile strength and the maximum tensile strength for whiskers possessing nearly equal diameter (5.1-5.4\,$\mu$m). Lines guide the eye.}
\end{center}
\end{figure}

Our predicted maximum tensile strength of Fe as a function of temperature $T$ [$\sigma_\text{m}(T)$] is shown in Fig.~\ref{fig:ITS} (solid line and circles). At 0\,K, the ITS amounts to 12.6\,GPa. For comparison, the measured room-temperature ultimate tensile strengths of bulk bcc iron ranges between 0.1\,GPa and 0.3\,GPa~\cite{Howatson:1972}, and the largest reported one is $\sim 6$\,GPa for $[001]$ oriented Fe whiskers~\cite{Brenner:1965}.
Temperature is found to have a severe impact on the ITS of Fe. Namely, $\sigma_\text{m}(T)$ remains nearly constant up to $\sim 350$\,K, decreases by $\sim 8\,\%$ between $\sim 350$\,K and $\sim 500$\,K compared to the 0\,K strength and then drops by $\sim 90\,\%$ between $\sim 500$\,K and $\sim 920$\,K. At temperatures above $\sim 1000$\,K, Fe can resist a maximum tensile strength of $\sim1.0\,$GPa. For all investigated temperatures, we found that the ideal Fe lattice maintains body-centered tetragonal (bct, lattice parameters $a$ and $c$) symmetry during the tension process and eventually fails by cleavage of the $(001)$ planes~\cite{inherent:property}.

\begin{figure}[bt]
\begin{center}
\resizebox{0.7\columnwidth}{!}{\includegraphics[clip]{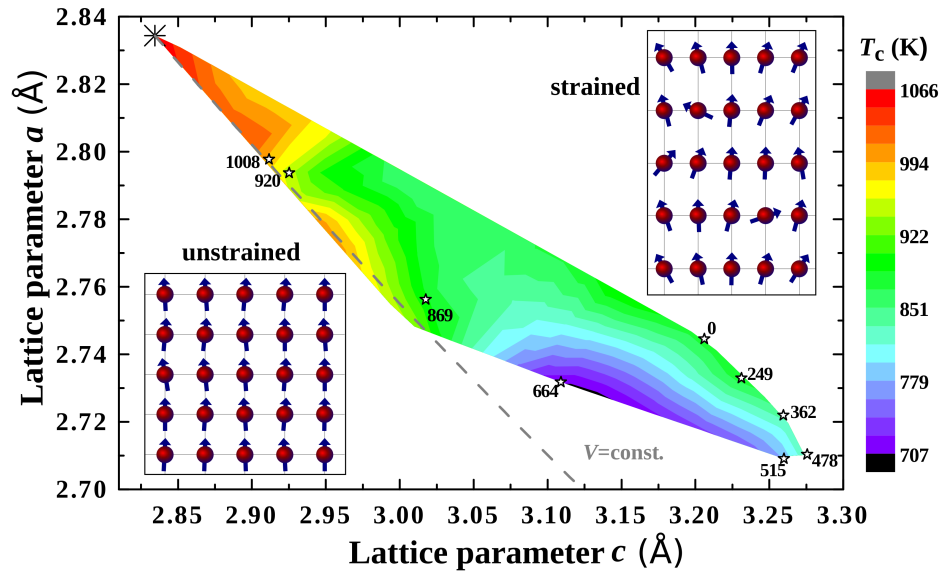}}
\caption{\label{fig:TC}Contour plot of $T_{\text{C}}$ of bct Fe as a function of the lattice parameters $a$ and $c$. The region where $T_\text{C}$ is shown is confined by the bcc ground state (black asterisk) and the failure points ($\epsilon_{\text{m}}$) at different $T$ (stars, numbers denoting $T$).  The Curie temperature corresponding to each failure point ($T_\text{C}^{\epsilon_{\text{m}}}$) can be read from the legend. The ground state bcc structure possesses the highest calculated $T_{\text{C}}$ in the region shown.
The dashed line represents the hyperbola of constant volume equal to the bcc equilibrium volume. The insets sketch local spins for the unstrained parent lattice and for the strained lattice at high temperature, illustrating the magnetic disorder increase with tensile strain.}
\end{center}
\end{figure}

 Figure~\ref{fig:ITS} also shows the ITS of Fe considering merely the thermal magnetic disorder effect (dashed line, open diamonds). Comparing these data to the full ITS (solid line and circles), it can be inferred that the main trend of $\sigma_{\text{m}}(T)$ is governed by the magnetic disorder term. The pronounced drop of the ITS at $\sim 500$\,K is related to the magnetic properties of strained Fe, more precisely to its magnetic transition temperature. The Curie temperature is shown as a function of the bct lattice parameters in Fig.~\ref{fig:TC}.
Compared to bcc Fe, $T_{\text{C}}$ of bct Fe is strongly reduced. The main trend is that $T_{\text{C}}$ decreases as the tetragonality ($c/a$) increases. A drop in $T_{\text{C}}$ with increasing $c/a$ was reported in previous fixed-volume calculations~\cite{Wang:2006}, which turns out to be the case for the present uniaxial tensile deformations as well.

In order to illustrate the significance of the lattice strain-induced reduction of $T_{\text{C}}$ on the strength of Fe, we computed an auxiliary ITS [$\sigma^c_\text{m}(T)$] assuming a constant $T_{\text{C}}$ for the strained lattice. We chose without loss of generality the present theoretical $T_{\text{C}}^{0}$ of bcc Fe (1066\,K) (see Table~S1 in the Supplementary information).
The magnetic disorder effect on $\sigma^c_\text{m}(T)$ is shown in Fig.~\ref{fig:ITS} (dotted line, open triangles). The drop of the ITS $\sigma^c_\text{m}(T)$ is shifted to higher temperatures compared to $\sigma_\text{m}(T)$. We conclude that the strongly reduced Curie temperatures of the distorted bct Fe lattices compared to bcc Fe are primarily responsible for the drop of the ITS at $\sim 500$\,K. In other words, lattice deformation destabilizes the magnetic order (shown schematically in insets of Fig.~\ref{fig:TC}) which leads to an unexpected softening of the lattice already at temperatures far below $T_{\text{C}}^{0}$.

\begin{figure}[hbtc]
\begin{center}
\resizebox{0.7\columnwidth}{!}{\includegraphics[clip]{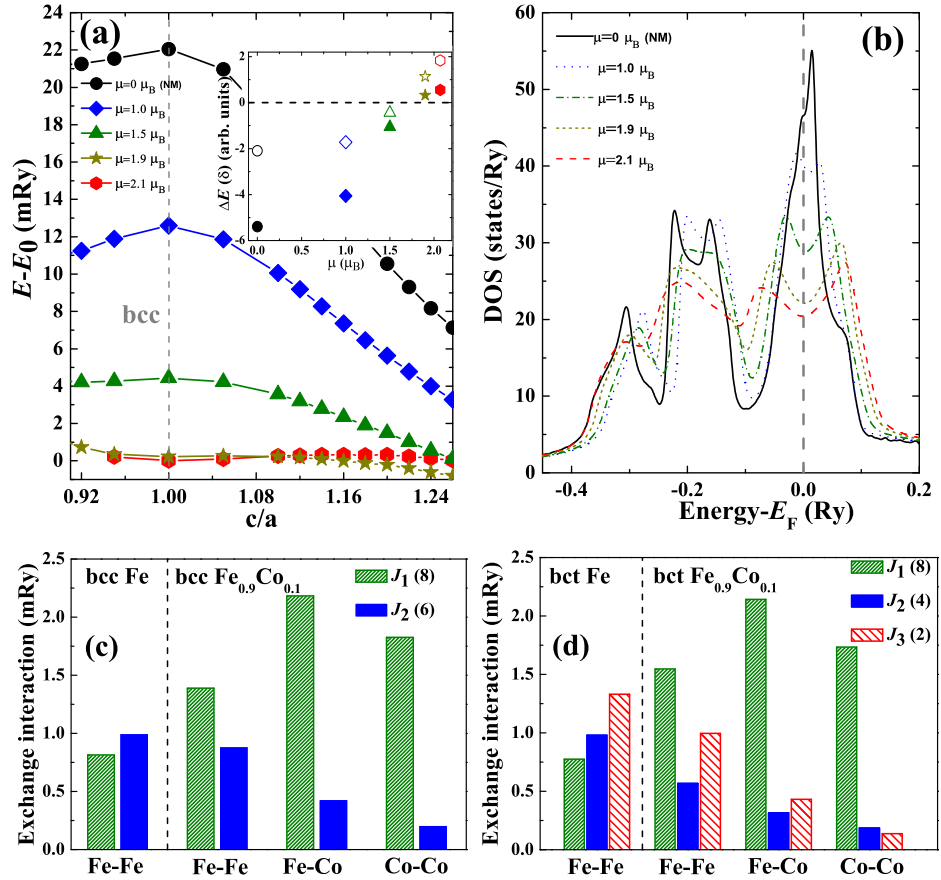}}

\caption{\label{fig:mechanism}(a) Total energy of NM bct Fe and PM bct Fe for various values of the local magnetic moment. The energies are plotted with respect to the energy of the PM bcc state with equilibrium local magnetic moment ($2.1\,\mu_{\text{B}}$), and the volume is fixed to that of PM bcc Fe. The inset displays the total energy change (open symbols) and the kinetic energy (solid symbols) change corresponding to a small volume-preserving tetragonal shear (arbitrary units). (b) Total DOS of PM bcc Fe for different local magnetic moments. (c) and (d) The exchange interactions $J_1$ and $J_2$ of bcc Fe and Fe$_{0.9}$Co$_{0.1}$ and $J_1$-$J_3$ of one representative bct structure ($a=2.732\,\textrm{\AA}$, $c=3.109\,\textrm{\AA}$).}
\end{center}
\end{figure}

 At temperatures $T\gtrsim 650$\,K, Fe reaches the maximum strength (fails) very close to the magnetic instability ($T/T_{\text{C}}^{\epsilon_{\text{m}}} > 0.9$). The strength of these magnetically barely ordered systems approaches the very low strength of the PM state. The emerging question is why the ideal tensile strength of PM Fe is so much lower than that of FM Fe.
We recall that bcc Fe is susceptible to the formation of sizable moments in both FM and PM states due to its peculiar electronic structure~\cite{Pettifor:1980,Gyorffy:1985}.
The nonmagnetic (NM) bcc Fe is thermodynamically and dynamically unstable as illustrated in Fig.~\ref{fig:mechanism}(a), where the total energy along the constant-volume Bain path is shown. The pronounced $E_{2g}$ peak in the NM density of states (DOS) at the Fermi level ($E_{\text{F}}$) [Fig.~\ref{fig:mechanism}(b)] drives the onset of FM order, that is, the majority and the minority spin channels populate and depopulate, respectively, causing an exchange split and moving the peak away from $E_{\text{F}}$~\cite{Mohn:2006}. When the FM equilibrium state is reached (at 0 K), $E_{\text{F}}$ is located near the bottom of the pseudo gap at approximately half band filling in the minority channel which explains the pronounced stability of FM bcc Fe. This is reflected by a deep minimum in the energy versus \emph{a} and \emph{c} map (see Fig.~S2 in the Supplementary information). The depth of this energy minimum in connection to the ITS is best characterized by the total energy cost to reach the ideal strain ($\epsilon_{\text{m}}$) from the unstrained bcc phase (strain energy), which amounts to $0.426$\,mRy/atom per $\%$\,strain in the FM state.

The formation of local magnetic moments in the PM state also stabilizes the bcc structure. This is illustrated in Fig.~\ref{fig:mechanism}(a), where we show the constant-volume Bain path of PM bct Fe for various fixed values of the local magnetic moment ($\mu$). It is found that increasing $\mu$ gradually stabilizes the bcc structure and for the equilibrium value of the local magnetic moment in the PM state ($2.1\,\mu_{\text{B}}$), a shallow minimum is formed on the tetragonal energy curve.
Figure~\ref{fig:mechanism}(a) (inset) displays the total energy change of the bcc structure corresponding to a small volume-preserving tetragonal deformation [$\Delta E(\delta) = E(\delta) - E(0)$] for the same $\mu$ values as in the main figure.
$\Delta E(\delta)$ is negative for NM Fe, but increasing $\mu$ turns $\Delta E(\delta)$ positive indicating the mechanical stabilization of PM bcc Fe upon PM moment formation. It is found that the mechanical stabilization due to local magnetic moment formation, embodied in the trend of $\Delta E(\delta)$ as a function of $\mu$, is in fact determined by the kinetic energy (solid symbols) and thus by the details of the electronic structure.
In the PM state, the local magnetic moment formation is due to a split-band mechanism for which the average exchange field is zero~\cite{Pettifor:1980,Gyorffy:1985}.
The disorder-broadened DOSs of PM Fe [Fig.~\ref{fig:mechanism}(b)] reveal that the formation of local magnetic moments effectively removes the peak at $E_{\text{F}}$ seen for NM Fe. However, this mechanism does not yield a comparably large cohesive energy increase as for the FM case.
The strain energy cost to reach $\epsilon_{\text{m}}$ starting from the shallow energy minimum equals $0.024$\,mRy/atom per $\%$\,strain in the PM phase, which is approximately 18 times smaller than in the FM case. Hence, the significantly lower ITS of PM Fe compared to that of FM Fe arises from the differences in the magnetism-driven stabilization mechanisms and the corresponding energy minima within the Bain configurational space (see Fig.~S2 in the Supplementary information).

The 0\,K ideal strength of Fe can be sensitively altered already by dilute alloying with, e.g., V or Co~\cite{Li:2014}. Apart from the intrinsic chemical effect of the solute atom, there is a magnetic effect due to the interaction between the solute atom and the Fe host.
Here we show that the strength of Fe can be significantly enhanced at high temperature by alloying with Co.

We assessed the magnetic properties of the random Fe$_{0.9}$Co$_{0.1}$ solid solution using the same methodology as for Fe.
The alloying effect of Co on $T_\text{C}$ is mainly connected to the strengthening of the first nearest neighbor exchange interactions ($J_1$)~\cite{Takahasi:2007,Lezaic:2007}.
We show the influence of Co on the eight first [$J_1(8)$] and six second [$J_2(6)$] nearest neighbor exchange interactions in Fig.~\ref{fig:mechanism}(c) for the bcc phase. $J_1$ between both similar and dissimilar atomic species increases significantly in Fe$_{0.9}$Co$_{0.1}$ compared to $J_1$ in pure Fe. At the same time, alloying with Co weakens $J_2$, but this effect is of lesser importance for $T_\text{C}$ than the strengthening of $J_1$.

We computed the ITS of the Fe$_{0.9}$Co$_{0.1}$ alloy at 0\,K~\cite{Li:2014} and in the high-temperature interval between 500\,K and 1000\,K. As shown in Fig.~\ref{fig:ITS}, Fe$_{0.9}$Co$_{0.1}$ exhibits $\sim 10$\,\% larger ITS than Fe at temperatures below $500$\,K. However, the strengthening impact of alloying on the ITS becomes more dramatic in the high-temperature region.
The physical origin underlying this effect is the higher Curie temperature of bct Fe$_{0.9}$Co$_{0.1}$ compared to bct Fe, which is a consequence of both the enhanced nearest neighbor exchange interactions and the stronger ferromagnetism in Fe$_{0.9}$Co$_{0.1}$, i.e.,
both the magnetic moment and exchange interactions are larger and more stable in the Fe-Co alloy than in pure Fe upon structural perturbation (tetragonalization)~\cite{Li:2014,Lezaic:2007,Acet:2010}.
The alloying effect of Co on $J_1(8)$, $J_2(4)$, and $J_3(2)$ (the two third nearest neighbor interactions) is shown in Fig.~\ref{fig:mechanism}(d) for one representative bct structure (with $a=2.732\,\textrm{\AA}$ and $c=3.109\,$\AA). The addition of Co leads to a strong increase of $J_1$ and a weakening of $J_2$ and $J_3$ similar to the bcc phase. For the chosen example, the $T_{\text{C}}$s of Fe and Fe$_{0.9}$Co$_{0.1}$ amount to 707\,K and 1053\,K, respectively, i.e., Co addition results in an increase of $T_{\text{C}}$ which is in magnitude much more pronounced for the bct structure than for the bcc one.

The predicted strong magneto-mechanical softening of Fe above $\sim$ 500\,K should be observable in flawless systems. Indeed, for single-crystalline Fe whiskers tensioned along [001]~\cite{Brenner:1956,Brenner:1958,Brenner:1965}, Brenner reported a pronounced temperature dependence of the average and maximum tensile strengths as shown in Fig.~\ref{fig:ITS} (inset). It seems plausible to assume that the maximum strength corresponds to whiskers with the lowest defect density and highest surface perfection. The failure of whiskers with diameter $<6\mu$m was reported to occur without appreciable plastic deformation. It is important to realize that the observed temperature gradient of the measured strength is much stronger than the one assuming only the thermally-activated nucleation of dislocations at local defects~\cite{Brenner:1965}. Hence, the observed softening of Fe whiskers requires additional intrinsic mechanisms emerging from the atomic-level interactions. The present results obtained for the tensile strength of ideal Fe crystal show a strong temperature gradient due to tensile-strain induced magnetic softening which actually nicely follows the experimental trend. The qualitative difference in the
theoretical and experimental temperature dependence may be ascribed to the low strain-rate inherent in experiment, which can allow for the activation of dislocation mechanisms~\cite{Fan:2012,Fan:2013}.

\section*{Conclusions}
 We have discovered a strong magneto-mechanical softening of iron single-crystals upon tensile loading. We showed that the strength along the $[001]$ direction persists up to $\sim$ 500 K, however, diminishes most strongly in the temperature interval $\sim$ 500 - 900\,K due to the loss of the net magnetization upon uniaxial strain. The strength in the paramagnetic phase is more than 10 times lower than that in the ferromagnetic phase. We found that Fe fails by cleavage at all investigated temperatures.

 We have demonstrated that the intrinsic strength of Fe at high temperatures is significantly enhanced by alloying with Co. This finding opens a way to carefully scrutinize the proposed connection~\cite{Zhu:2010,Yue:2011} between the tensile strength of a defect-free ideal crystal and the measured tensile strength of single-crystal whiskers or nanoscale systems (e.g., nanopillars). Measuring the tensile strength of Fe-Co whiskers at elevated temperature could verify the here predicted differences in the high-temperature intrinsic mechanical properties of Fe and Fe-Co alloy.

Extending the above mechanism to other systems, we expect that the large magnetic effect on the strength exists for all dilute Fe alloys and that the magneto-mechanical softening temperature can be sensitively tuned by proper selection of the alloying elements. Solutes that strengthen (weaken) the ferromagnetic order in the tetragonally distorted Fe increase (decrease) the intrinsic strength of defect-free Fe crystals at temperatures above (below) $\sim$ 600\,K.

\section*{Method}
The adopted first-principles method is based on density-functional theory as implemented in the exact muffin-tin orbitals (EMTO) method~\cite{EMTO:1,EMTO:2,EMTO:3} with exchange-correlation parameterized by Perdew, Burke, and Ernzerhof (PBE)~\cite{PBE,Perdew:1996E}. EMTO features the coherent-potential approximation (CPA), which allows to describe the disordered PM state by the disordered local moment (DLM) approach~\cite{Gyorffy:1985}.

The ITS ($\sigma_\text{m}$) with corresponding strain ($\epsilon_{\text{m}}$) is the first maximum of the stress-strain curve, $\sigma(\epsilon)=\frac{1+\epsilon}{V(\epsilon)}\frac{\partial G}{\partial \epsilon}$ [$V(\epsilon)$ and $G(\epsilon)$ are the relaxed volume and the calculated free energy at strain $\epsilon$], upon uniaxial loading.
The tensile stress was determined by incrementally straining the crystal along the $[001]$ direction (the weakest direction for bcc crystals) and taking the derivative of the free energy [$G(\epsilon)$] with respect to $\epsilon$.
The two lattice vectors perpendicular to the $[001]$ direction were relaxed at each value of the strain allowing for a possible symmetry lowering deformation with respect to the initial body-centered tetragonal symmetry~\cite{Li:2014}.

The temperature induced contribution from electronic excitations to the ITS was considered by smearing the density of states with the Fermi-Dirac distribution~\cite{Wildberger:1995}. To measure the effect of explicit lattice vibrations, we computed the vibrational free energy as a function of strain in the vicinity of the stress maximum within the Debye model~\cite{Grimvall:1999} employing an effective Debye temperature which was determined from bulk parameters~\cite{Moruzzi:1988}.

In order to take into account the effect of thermal expansion, the ground-state bcc lattice parameter at temperature $T$ was obtained by rescaling the theoretical equilibrium lattice parameter with the experimentally determined thermal expansion~\cite{Basinski:1955,Nix:1941}.
The expanded volume was stabilized by a hydrostatic stress $p_T$ derived from the partial derivative of the bcc total energy with respect to the volume taken at the expanded lattice parameter corresponding to $T$.
$p_T$ was accounted for in the relaxation during straining the lattice. To this end, the relaxed lattice maintained an isotropic normal stress $p_T$ in the $(001)$ plane for each value of the strain.

We modeled the effect of thermal magnetic disorder on the total energy by means of the partially disordered local moment
(PDLM) approximation~\cite{Khmelevskyi:2003,Razumovskiy:2011b}. Within PDLM, the magnetic state of Fe and Fe$_{0.9}$Co$_{0.1}$ alloy are described as a binary Fe$^\uparrow_{1-x}$Fe$^\downarrow_{x}$ and a quaternary (Fe$_{0.9}^\uparrow$Co$_{0.1}^\uparrow$)$_{1-x}$(Fe$_{0.9}^\downarrow$Co$_{0.1}^\downarrow$)$_{x}$ alloy with concentration $x$ varying from $0$ to $0.5$ and anti-parallel spin orientation of the two alloy components. Case $x=0$ corresponds to the completely ordered FM state with magnetization $m=1$. As $x$ is gradually increased to $0.5$, the magnetically random PM state lacking magnetic long and short range order ($m=0$) is obtained (DLM state).
The PDLM approach describes the energetics underlying the loss of magnetic order. Connection to the temperature is provided through the magnetization curve [$m(\tau)$, $\tau\equiv T/T_{\text{C}}$ being the reduced temperature] which maps the computed $m$ to $T$. It is important to realize that for Fe and Fe$_{0.9}$Co$_{0.1}$ the thermal spin dynamics embodied in the shape of $m(\tau)$ and in the $T_{\text{C}}$ value change under the presently applied uniaxial loading.
The computational details for $T_{\text{C}}$ and $m(\tau)$ are given in the Supplementary information.


\section*{Acknowledgements}
We thank E. K. Delczeg-Czirjak, M.\,D.\,Kuz'min, and L.\,Udvardi for helpful discussions.
The Swedish Research Council, the Swedish Foundation for Strategic Research,the European Research Council, the China Scholarship Council and the Hungarian Scientific Research Fund (research project OTKA 84078 and 109570), and the National Magnetic Confinement Fusion Program of China (2015GB118001) are acknowledged for financial support. L.B. acknowledges financial support from the Swedish e-Science Research Centre (SeRC) and G\"oran Gustafsson Foundation.  The computations were performed using resources provided by the Swedish National Infrastructure for Computing (SNIC) at the National Supercomputer Centre in Link\"oping. L.S. was supported by the European Union, cofinanced by the European Social Fund, in the framework of the T\'{A}MOP 4.2.4.A/2-11-1-2012-0001 National Excellence Program.

\section*{Author contributions statement}

 L.V., J.Z., S. S., and X. L. designed research.  X. L. and S. S. performed research and analyzed the results. E.S, L.B., H.Z., and L.S. contributed to the calculations.   X. L., S. S., and L.V. wrote the paper.  All authors reviewed the paper.

 \section*{Additional information}
 \textbf{Competing financial interests} The authors declare no competing financial interest.

\end{document}